# Formation of white-eye pattern with microdischarge in an air dielectric barrier discharge system

Yafeng He, Lifang Dong*, Weili Liu, Hongfang Wang, Zengchao Zhao, and Weili Fan

*College of Physics Science & Technology, Hebei University, Baoding, 071002, China*

We report on the first observation of white-eye pattern in an air dielectric barrier discharge. The patterned discharges undergo a development as following: random spots – quasihexagonal pattern – hexagonal pattern (type I) – hexagonal pattern (type II) – white-eye pattern – chaos as the voltage is increased. The spatiotemporal characteristics of patterned discharges are investigated by using an optical method. Results show that the two discharge modes, uniform mode and filamentary mode, are actually two different spatial presentations of the same origin: the microdischarge. From the viewpoint of pattern dynamics, the white-eye pattern results from a 3-wave resonance interaction.

Dielectric barrier discharge (DBD, also referred to as silent discharge) is a type of non-equilibrium, non-thermal gas discharge, which has given rise to the two subjects of great interest at present, i.e., investigation of pattern formation in DBD and realization of atmospheric pressure glow discharge. It is characterized by a dielectric between electrodes and governed by the electron avalanches ignited in the gap. Based on the evolution of an avalanche, DBD can exhibit two major modes: homogeneous or glow mode and filamentary mode [1, 2]. In the homogeneous mode, numerous small avalanches develop under low field avoiding streamer formation. The discharge shows a homogeneous feature across the electrodes and has only one current pulse per half cycle of the applied voltage [3]. There is also another type of discharge with a homogeneous appearance (referred to as streamer-coupling discharge) obtained under specific conditions, in which the preionization associated with a rapid increase of the voltage efficiently induces the simultaneous development of so many streamers that they interfere and develop as a single discharge channel covering a large area [1]. In the filamentary mode, a number of individual breakdown channels, which are referred to as microdischarges, occur attributed to the secondary avalanches and high internal field of primary avalanches. The filamentary mode is characterized by numerous microdischarges distributed in the discharge gap and by many current pulses per half cycle of voltage [3]. In fact, all the discharges in different modes are composed of plasma domains induced by the avalanches. Thus, these plasma domains must be related to the observed patterns in DBD. Boyers reported that the hexagonal pattern, obtained in atmospheric pressure helium, is an assembly of glow discharge plasma domains [4]. In one review article, Kogelschatz showed that the 2-D discharge pattern obtained under certain conditions is presumably an arrangement of a number of smaller diffuse discharge areas [5]. Purwins considered the discharge pattern as a regular arrangement of filaments through interaction with each other [6]. But there is still no direct experimental evidence indicating the composition of the discharge patterns.

In this letter, we report on an experimental study of the patterned discharges in an air dielectric barrier discharge. A white-eye pattern is observed for the first time in DBD. The spatiotemporal characteristics of patterned discharges are investigated. It is found that the two discharge modes, uniform mode and filamentary mode, are actually two different spatial presentations of the same origin: the microdischarge.

The experimental device used in this work is similar to that described previously [7, 8]. Two cylindrical containers, with inner diameter of 70 mm, sealed with 1.5 mm thick glass plates, are filled with water. Here, the water acts as a liquid electrode. The parallel glass serves as the dielectric layers. A circular glass ring with the diameter of 70 mm serving as a boundary, limits the lateral diffusion of discharge gas. The electrode system is situated inside a vacuum chamber filled with ambient air. The gas pressure is varied from 76 to 760 Torr, and the gas gap can be changed from 0.1 mm to 4 mm. The gas is refreshed when changing the gas pressure or the gas gap width.

The temperature of water electrodes increases no more than 5℃ during an experiment (within about 10 minutes). Light emission of the discharge is detected by PMTs (photoelectric multiple tube, RCA7265) and recorded with an oscilloscope (Tektronix TDS3054B). A CCD camera is used to record the images of discharge. The visible emission spectrum from the discharge is obtained through a monochromator (Acton Spectra Pro 2750, grating: 300 grooves/mm).

Figure 1 shows the development of patterns: random spots – quasihexagonal pattern – hexagonal pattern (type I) – hexagonal pattern (type II) – white-eye pattern – chaos as the voltage was increased. All these patterns consist of spots and an illuminated background. It can be seen that the characteristics of the spots in Fig.1 a-c are different from that in Fig.1d-f. We refer to the spots with large area and low intensity in Fig.1 a-c as type I spots, and the spots with small area and high intensity in Fig.1 d-f as type II spots. In fact, there is a spontaneous jumping transition between the two types of spots with increasing voltage. The transition can happen to and fro when the voltage reaches 2.9 kV with other parameters fixed. The mean interparticle space could hardly change, and the system exhibits hexagonal pattern, while the background is not intense enough to form the white-eye pattern. However, the luminance of the illuminant background increases with the voltage. When the voltage reaches about 3.2 kV, the background can be seen clearly. At this time, the background forms a honeycomb pattern surrounding a hexagonal lattice (type II spots), and the white-eye pattern comes into being as shown in Fig.2. So, the white-eye pattern is composed of the hexagonal lattice of type II spots and the intense background.

It has been seen that the background, type I spots and type II spots have different spatial characteristics, but there is still no idea whether they belong to different discharge modes. Therefore, it is necessary to investigate their spatiotemporal characteristics for understanding the pattern formation. Here, we employ an optical method with PMTs to study the temporal and spatial characteristics of spots and background in patterns. The spots in Fig.1 a, f are drift at a velocity of the order of mm/s, while they are nearly stable in Fig.1 b-e. So it is available to target and track certain spot with movable PMTs. For the temporal aspect, the time resolution can be realized at the ns scale, which is less than the duration of discharge in air. For the spatial aspect, the size of spots in Fig.1 is larger than 0.4 mm. We use a lens to image the pattern and a hole-aperture to enable part of the light of discharge pattern, such as partial spot, to enter the PMT. All the measured areas of the pattern are of the order of $10^{-2}mm^{-2}$, which is about 5% of area of one spot. So the spatial resolution can be realized at the micron scale. Therefore, this optical method is available for the high spatiotemporal resolved measurement of patterned discharge.

In the experiment, the signals of applied voltage, current, total light emission, part of background, part of spot, and a whole spot together with partial background nearby are measured with increasing voltage. It is found that the background, type I spots and type II spots have distinctive characteristics. Typical results are shown in Fig.3, which corresponds to a hexagonal pattern (Fig.1 c) and the white-eye pattern (Fig.1 e). In the following content we'll discuss the results of measurement of the background, type I spots and type II spots.

As can be seen in Fig.3 a (iv), b (iv), the discharges of partial background under different voltages have similar signals with discontinuous and random current pulses in consecutive half cycles. This implies that there should be a certain microdischarge channel igniting randomly or moving through the measured area within several half cycles, and the observed uniform background is an integration of a large number of microdischarges. Accordingly, the background operates at uniform discharge that has uniform appearance and narrow current pulses.

The spatiotemporal characteristics of partial type I (type II) spots are similar as shown in Fig.3 a (v) (Fig.3 b (v)), but distinctive between type I spots and the type II ones. As can be seen from Fig.3 a(v), the partial spot (type I) does not discharge at every cycle, which likes the case of background discharge. So the hexagonal pattern in Fig.1 c (type I spots) actually is an arrangement of a number of smaller uniform discharge areas. From Fig.3 b(v), we found that the partial spot (type II) measured within area of $10^{-2}mm^{-2}$ does discharge in consecutive half cycles. Thus the hexagonal lattice of white-eye pattern in Fig.1 e consists of filaments (type II spots) that present the microdischarges ignited on the same position in consecutive half cycles.

A whole type I (type II) spot together with partial background are measured simultaneously, and results are shown in Fig.3 a(vi) (Fig.3 b(vi)). It is found that a whole spot of type I discharges at every half cycle, although part of that discharges discontinuously in consecutive half cycles (Fig.3 a(v)). The type I spot further proved to be operated at uniform discharge.

From the above results, we can draw a conclusion that the background and the type I spots operate as a uniform discharge which is an integration of a large number of microdischarges, while the type II spots are actually filaments that present the microdischarges ignited on the same position in consecutive half cycles. They are the different spatial presentations of microdischarges.

Furthermore, the optical emission spectroscopy of the white-eye pattern is investigated. As Choi stated, the relative intensity of the first negative band system of $N_2^+$ (391.4nm) and atomic oxygen (777.1nm) can be used as a criterion to determine the discharge mode [9]. In the present experiment, we obtain the relative intensity of the first negative band system of $N_2^+$ (391.1nm) and $N_2$ (380.4nm) after the normalization of the second positive band system of $N_2$ (337.1nm). The relative intensity of background (type II spots) is about 3.9 (3.6). In addition, by analyzing the vibrational temperature of white-eye pattern from the second positive band system of $N_2$, it is found that the vibrational temperature of the background (type II spots) is about 2787k (2767k). These results show that the relative intensity of $I_{N_2^+}/I_{N_2^*}$ and the vibrational temperature of the background and the type II spot are nearly equal respectively. So we can come to the same conclusion that the background and the type

II spots are two different spatial presentations of microdischarge.

As far as the above patterned discharge is concerned, two particular aspects had to be pointed out. On the one hand, the separation of spot and background in space gives rise to two current pulses humps at every half cycle. The first (second) one only contains the discharges of spots (background), which can be seen in Fig.3. In general, once the voltage reaches a critical one, the spots discharge and can self-organize into patterns under proper control parameters. There exist some places between spots where the effective electric field is higher than that in the spot due to the action of surface charges. A continuous rise of voltage on rising edge can lead to the discharge of background among spots, which refers to as interpolative discharge. Furthermore, in the air discharge, oxygen is electronegative, it captures electrons and weakens the electron avalanche processes around the filaments. This reduces the diameter of filament and increase the ignition voltage of the gas around filaments, which leads to the dark ring round filaments (type II spot) in white-eye pattern.

On the other hand, the transition between the two types of spots is a jumping process when the voltage reaches a critical value. The type I spot oscillates within a large spatial scale leading to the uniform spatial presentations of microdischarges. The interaction among spots is enhanced as the voltage is increased. Then the spots are further restricted close to their balance places, resulting in the formation of type II spots when the voltage reaches a critical value. More detailed experiments are needed in the coming work in order to explain this transition.

From the viewpoint of dynamics of pattern formation, the white-eye pattern in reaction-diffusion system has proved an interaction between two modes with different wavelengths [10]. To study the resonance of white-eye pattern in Fig.1, we analyzed the Fourier spectra of Fig.1c and e. As is well known, hexagonal pattern satisfies a resonance condition $\vec{q}_1 = -\vec{q}_3 - \vec{q}_2$ as shown in Fig.4 a. With increasing voltage, the background is intensified gradually and white-eye pattern forms following the hexagonal pattern. Meanwhile, a mode (k) induced by the background is excited and interacts with the q mode. A new resonance relationship appears in white-eye pattern, which satisfies $\vec{k}_1 = \vec{q}_1 - \vec{q}_2$ (Fig.4 b). So the white-eye pattern in DBD system results from a 3-wave resonance.

In conclusion, two discharge modes, uniform and filamentary, are observed in the patterned discharges in air. They are actually the different spatial presentations of the same origin: microdischarge. The observed white-eye pattern for the first time consists of honeycomb background and hexagonal lattice which have proved to be operated at uniform and filamentary mode respectively. From the viewpoint of pattern dynamics, the white-eye pattern results from a 3-wave resonance interaction.

This work is supported by the National Natural Science Foundation of China under Grants No.10575027 and No.10375015，the Specialized Research Fund for the Doctoral Program of higher Education of China (Grant No. 20050075001), and the Natural Science Foundation of Hebei Province, China (Grants No. A2006000950 and No. A2004000086).

*Corresponding author: Donglf@mail.hbu.edu.cn

**Figure Captions:**

FIG. 1. Evolution of patterns with increasing voltage. (a) random spots, 2.3 kV; (b) quasihexagonal pattern, 2.7 kV; (c) hexagonal pattern (type I spot), 2.9 kV; (d) hexagonal pattern (type II spot) , 2.9 kV; (e) white-eye pattern, 3.4 kV; (f) chaos, 4.3 kV. The other control parameters: $p$=0.12 atm, $d$=2 mm, $f$=50 kHz.

FIG. 2. White-eye pattern. (a) original image. (b) partial image after enhancing the contrast. The parameters are the same as those in Fig.1 e.

FIG. 3. Discharge waveforms (a, b) of hexagonal pattern (Fig.1 c) and white-eye pattern (Fig.1 e) respectively. The signals i-vi correspond to the applied voltage, current, total light emission, part of background, part of spot, and a whole spot together with partial background nearby respectively. The measured area of pattern is of the order of $10^{-2}$mm$^{-2}$ which is kept about 5% of that of one spot.

FIG. 4. Spatial spectra of hexagonal pattern (a) and white-eye pattern (b).

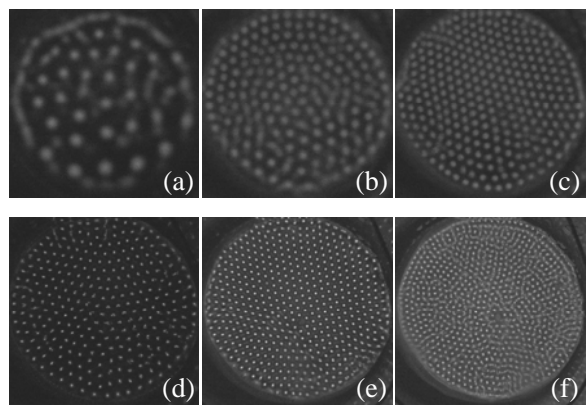

Figure 1

Yafeng He

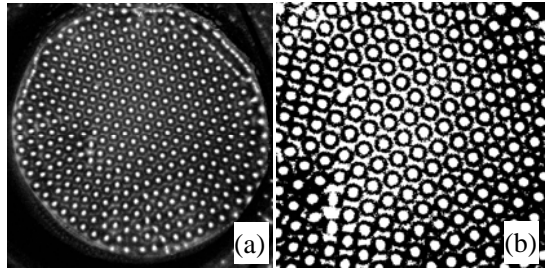

Figure 2

Yafeng He

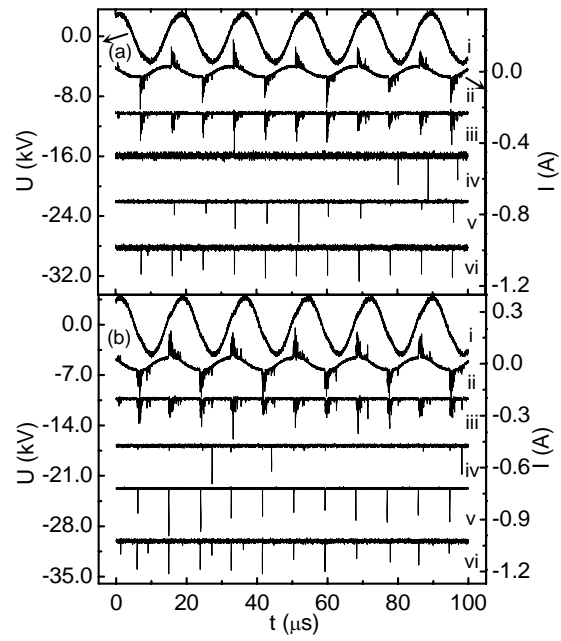

Figure 3

Yafeng He

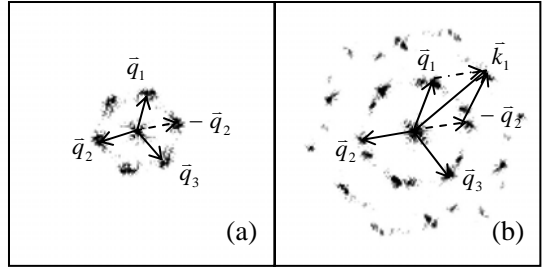

Figure 4

Yafeng He